\documentclass[aps,prl, amsmath, showpacs, preprintnumbers,superscriptaddress, twocolumn,sort&compress,floatfix, amssymb]{revtex4}
\usepackage{graphicx}
\usepackage{rotating}
\usepackage{dcolumn}
\usepackage{bm}
\usepackage{color}
\usepackage{mathptmx, textcomp}
\usepackage[latin1]{inputenc}
\usepackage{braket}

\begin{document}

\author{Elmar Haller}
\author{Manfred J. Mark}
\author{Russell Hart}
\author{Johann G. Danzl}
\author{Lukas Reichs\"ollner}\affiliation{Institut f\"ur Experimentalphysik and Zentrum f\"ur Quantenphysik, Universit\"at Innsbruck, Technikerstra{\ss}e 25, 6020 Innsbruck, Austria}
\author{Vladimir Melezhik}
\affiliation{Bogoliubov Laboratory of Theoretical Physics, Joint Institute for Nuclear Research,
Dubna, 141980 Dubna, Russia}
\author{Peter Schmelcher}
\affiliation{Zentrum f\"ur Optische Quantentechnologien, Universit\"at Hamburg, Luruper Chaussee 149, 22761 Hamburg, Germany}
\author{Hanns-Christoph N\"agerl}\affiliation{Institut f\"ur Experimentalphysik and Zentrum f\"ur Quantenphysik, Universit\"at Innsbruck, Technikerstra{\ss}e 25, 6020 Innsbruck, Austria}

\title{Confinement-Induced Resonances in Low-Dimensional Quantum Systems}

\date{\today}

\pacs{34.50.-s, 03.65.Nk, 05.30.Jp, 37.10.Jk}

\begin{abstract}
We report on the observation of confinement-induced resonances in strongly interacting quantum-gas systems with tunable interactions for one- and two-dimensional geometry. Atom-atom scattering is substantially modified when the s-wave scattering length approaches the length scale associated with the tight transversal confinement, leading to characteristic loss and heating signatures. Upon introducing an anisotropy for the transversal confinement we observe a splitting of the confinement-induced resonance. With increasing anisotropy additional resonances appear. In the limit of a two-dimensional system we find that one resonance persists.
\end{abstract}

\maketitle

Low-dimensional systems have recently become experimentally accessible in the context  of ultracold quantum gases. For a two-dimensional (2D) geometry, the Berezinskii-Kosterlitz-Thouless (BKT) transition has been observed \cite{Hadzibabic2006}, and in one dimension the strongly-correlated Tonks-Girardeau (TG) \cite{Girardeau1960, Kinoshita2004,Paredes2004,Syassen2008,Haller2009} and super-Tonks-Girardeau (sTG) gases \cite{Haller2009} have been realized. In these experiments steep optical potentials freeze out particle motion along one or two directions and restrict the dynamics to a plane or to a line. Such quasi-2D or quasi-1D systems can be realized with ultracold gases when the kinetic and the interaction energy of the particles are insufficient to transfer the particles to transversally excited energy levels. Whereas the confinement removes motional degrees of freedom, it also provides an additional structure of discrete energy levels that can be used to modify scattering along the unconfined direction and by this to effectively control the interaction properties of the low-dimensional system \cite{Olshanii1998,Petrov2000,Bergeman2003}. In this Letter, we investigate the few-body scattering processes that give rise to the capability to tune interactions and hence to drastically alter the properties of low-dimensional many-body quantum systems \cite{Haller2009}.

In three-dimensional (3D) geometry magnetically-induced Feshbach resonances (FBRs) \cite{Chin2009} allow tuning of the inter-particle interaction strength. A FBR occurs when the scattering state of two atoms is allowed to couple to a bound molecular state. Typically, scattering state and bound state are brought into degeneracy by means of the magnetically tunable Zeeman interactions. For particles in 1D and 2D geometry a novel type of scattering resonance occurs. Coupling between the incident channel of two incoming particles and a transversally excited molecular bound state generates a so-called confinement-induced resonance (CIR) \cite{Olshanii1998,Petrov2000,Bergeman2003,LowDSystems,Kim2005,Naidon2007}. A CIR occurs when the 3D scattering length $a_{\text{3D}}$ approaches the length scale that characterizes the transversal confinement, i.e. the harmonic oscillator length $a_{\perp} = \sqrt{\hbar /(m \omega_{\perp})}$ for a particle with mass $m$ and transversal trapping frequency $\omega_{\perp}$. This causes the 1D coupling para\-meter $g_\mathrm{1D} = \frac{2\hbar^2 a_{\text{3D}}}{m a_{\perp}^2} \frac{1}{1-C a_{\text{3D}}/a_{\perp}}$ to diverge at $a_{\perp} = C a_{\text{3D}}$, where $C=1.0326$ is a constant \cite{Olshanii1998, Bergeman2003}. The CIR allows tuning of interactions from strongly repulsive to strongly attractive and thus represents a crucial ingredient for the control of interactions in a low-dimensional system. Modification of scattering properties due to confinement has been measured near a FBR for fermions \cite{Guenter2005}, and, recently, a CIR has been observed for a strongly-interacting 1D quantum gas of bosonic Cs atoms and was used to drive the crossover from a TG gas with strongly repulsive interactions to an sTG gas with strongly attractive interactions \cite{Haller2009}. Here, for an ultracold quantum gas of Cs atoms with tunable interactions, we study the properties of CIRs by measuring particle loss and heating rate and, in particular, confirm the resonance condition $a_{\perp} = C a_{\text{3D}}$ for symmetric 1D confinement. For the case of transversally anisotropic confinement we find that the CIR splits and, to our surprise, persists for positive $a_{\text{3D}}$ even when the anisotropy reaches the limit of a 2D system.

\begin{figure}[t]
 \includegraphics[width=8.5cm] {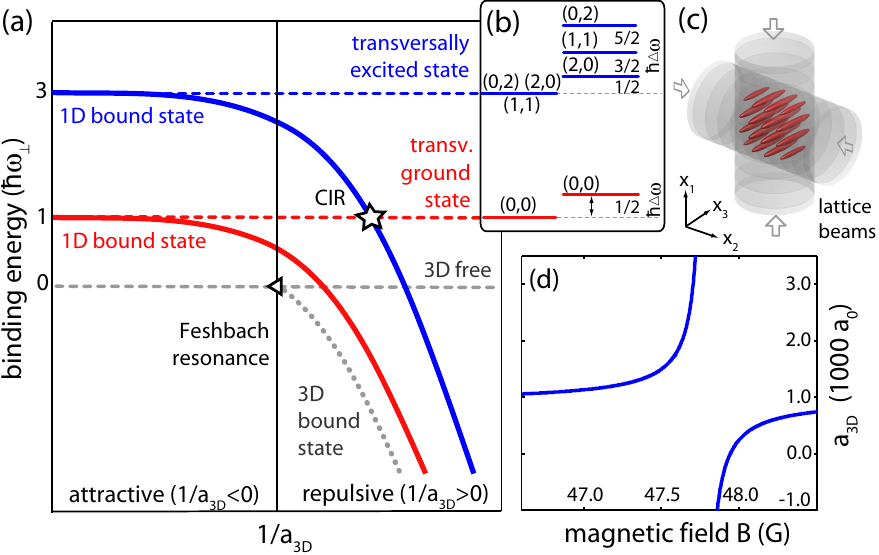}
 \caption{(color online) (a) Illustration of the mechanism responsible for a CIR, see Ref.\cite{Bergeman2003} and text for details. The energy levels near a scattering resonance are plotted as a function of $1/a_{\text{3D}}$. The CIR occurs for $C a_{\text{3D}}=a_\perp$ when scattering atoms are allowed to couple to transversally excited bound states. (b) indicates the shift and splitting for anisotropic confinement characterized by $\Delta \omega$. (c) Experimental configuration. Two laser beams create an optical lattice that confines the atoms to an array of approximately $3000$ independent, horizontally-oriented elongated 1D tubes. (d) Tuning of $a_{\text{3D}}$ is achieved by means of a FBR with a pole at $B=47.78(1)$ G \cite{Lange2009}.} \label{fig1}
\end{figure}

Figure~\ref{fig1}(a) reviews the basic mechanism that causes a CIR for zero collisional energy in 1D \cite{Bergeman2003}. It is assumed that in 3D the scattering potential supports a single universal bound state for strong repulsive interactions (dotted line) \cite{Chin2009}. The point where the incoming channel of two colliding atoms and the universal dimer state are degenerate marks the position of a 3D FBR (triangle). In 1D, strong transversal confinement shifts the zero-energy of the incoming channel (middle dashed line) and introduces a transversally excited state (upper dashed line). As a result of the strong confinement, which modifies the long-range part of the molecular potential, the universal dimer state with binding energy $E_\mathrm{B}$ (lower solid line) exists also for attractive interactions \cite{Moritz2003} whereas the original 3D FBR has disappeared. Instead, there is a CIR (star) when the incoming scattering channel becomes degenerate with the transversally excited molecular bound state (upper solid line). It is assumed that the binding energy of this state is also $E_\mathrm{B}$, shifted by $2\hbar\omega_\perp$\cite{Olshanii1998}. In more detail, as depicted in Fig.~\ref{fig1}(b), we assume that the energy levels of non-interacting atoms, as a result of cylindrically symmetric transversal confinement, can be approximated by those of a 2D harmonic oscillator with $E_{n_1,n_2} = \hbar \omega_{\perp}(n_1 + n_2 + 1)$ and quantum numbers $n_1$ and $n_2$ belonging to the two Cartesian directions. Scattering atoms \cite{Separation} in the transversal ground state $(0,0)$ can couple to the excited states $(n_1, n_2)$ if the parity of the total wave function is preserved \cite{Kim2005}. The energetically lowest allowed excited states are threefold degenerate with an energy $E=3 \hbar\omega_{\perp}$ and with quantum numbers $(1,1)$, $(2,0)$ and $(0,2)$. For the transversally symmetric confinement, they contribute towards a single CIR \cite{Bergeman2003}. However, the contribution of the state (1,1) is negligible due to the zero contact probability of the atoms and the short-range character of the interatomic interaction. Unequal transversal trapping frequencies $\omega_1$ and $\omega_2 = \omega_1 + \Delta \omega$ lift this degeneracy and shift the energy levels according to $ E_{n_1,n_2} = \hbar \omega_1(n_1 + n_2 + 1) + \hbar \Delta \omega (n_2+ 1/2)$. One thus expects a splitting of the CIR.

We start from a tunable Bose-Einstein condensate (BEC) of $1.0$ to $1.4\times10^5$ Cs atoms in the energetically lowest hyperfine sublevel \cite{Kraemer2004} confined in a crossed-beam optical dipole trap and levitated against gravity by a magnetic field gradient of $|\nabla B| \approx 31.1$ G/cm. Tunability of $a_{\text{3D}}$ is given by a FBR as shown in Fig.~\ref{fig1}(d) with its pole at $B_0=47.78(1)$ G and a width of $164$ mG \cite{Kraemer2004,Lange2009}. The resonance resides on top of a slowly varying background that allows tuning of $a_{\text{3D}}$ from $0$ to values of about $1000 \ a_0$, where $a_0$ is Bohr's radius. Using the FBR, we can further tune to values up to $a_{\text{3D}} \approx 6000 \ a_0$ given our magnetic field control with an uncertainty of $\Delta B \approx 10$ mG. We convert $B$ into $a_{\text{3D}}$ using the FBR parameters from Ref.\cite{Lange2009}. The BEC is produced at $a_{\text{3D}} \approx 290 \ a_0$. We load the atoms within $300$ ms into an optical lattice, which is formed by two retro-reflected laser beams at a wavelength of $\lambda=1064.49(1)$ nm and with a beam waist of approximately $350 \ \mu$m, one propagating vertically and one propagating horizontally as illustrated in Fig.~\ref{fig1}(c). These lattice beams confine the atoms to an array of approximately $3000$ horizontally oriented, elongated 1D tubes with a maximum occupation of $60$ atoms at a linear peak density of approximately $n_\text{1D} \approx 2 / \mu$m. Weak longitudinal confinement results from the Gaussian-shaped intensity distribution of the beams. We raise the lattice to a depth of typically $V = 30 \ E_R$, where $E_R=h^{2}/(2 m \lambda^{2})$ is the photon recoil energy. At this depth, the resulting transversal and longitudinal trap frequencies are $\omega_\perp = 2 \pi \times 14.5$ kHz and $ \omega_\parallel = 2 \pi \times 16$ Hz and we then have $a_{\perp} \approx 1370 \ a_0$. After loading we slowly ramp down $|\nabla B|$ in $50$ ms and adiabatically increase $a_{\text{3D}}$ to $915$ a$_0$ in $100$ ms to create a TG gas with well-defined starting conditions near the CIR \cite{Haller2009}. To detect the CIR as a function of $B$, manifested by a loss resonance, we quickly set $B$ in less than $200 \ \mu$s to the desired value, wait for a hold time of typically $\tau=200$ ms, and then measure the number $N$ of remaining atoms by absorption imaging. For this, we re-levitate the atoms, ramp down the lattice beams adiabatically with respect to the lattice band structure, and allow for $50$ ms of levitated expansion and $2$ ms time-of-flight. Note that $\tau$ is chosen to be much longer than the lifetime of the sTG phase \cite{Haller2009}.

\begin{figure}[t]
\includegraphics[width=8.5cm] {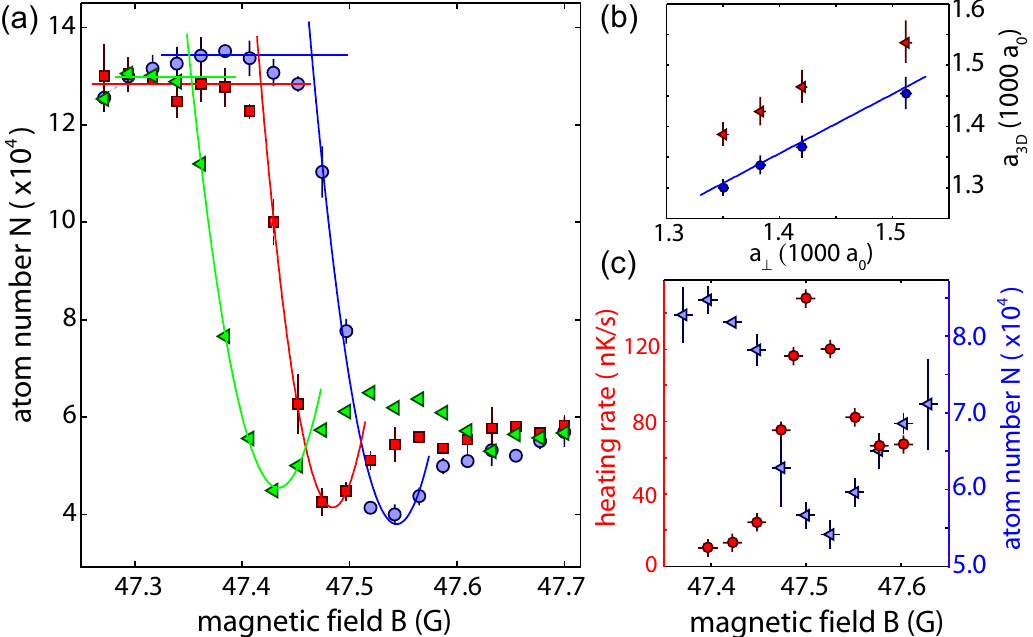}
 \caption{(color online) Particle loss and heating rates in the vicinity of a CIR. (a) The number $N$ of remaining atoms after $\tau=200$ ms shows a distinct drop (``edge'') when $B$ is scanned across the CIR. A clear shift of the position of the edge to lower values for $B$ can be observed when the transversal confinement is stiffened, $\omega_{\perp} = 2 \pi \times  (0.84,0.95,1.05)\times 14.2(2)$ kHz (circles, squares, triangles). (b) Position of the edge (circles) as determined from the intersection point of a second-order polynomial fit to the minimum for $N$ and the initial horizontal baseline as shown in (a), converted into values for $a_{\text{3D}}$. For comparison, the position of the minimum (triangles) is also shown. The solid line is given by $C a_{\text{3D}} = a_\perp$ with the predicted value $C=1.0326$. (c) Heating rates near the CIR (circles). For comparison, $N$ is also shown (triangles). For this measurement, $\omega_{\perp}=2\pi \times 12.0(2)$ kHz. All error bars reflect $1\sigma$ statistical uncertainty.} \label{fig2}
\end{figure}

We observe the CIR in the form of an atomic loss signature as shown in Fig.~\ref{fig2}. We attribute the loss near the resonance to inelastic three-body \cite{Weber2003} or higher-order collisions \cite{Ferlaino2009}, which lead to molecule formation and convert binding energy into kinetic energy, causing trap loss and heating, similar to the processes observed near a FBR \cite{Chin2009}. In our case, inelastic two-body processes can be ruled out for energetic reasons and single-particle loss occurs on the timescale of tens of seconds. In Fig.~\ref{fig2}(a) the CIR can be identified as a distinct ``edge'' for the atom number $N$. Initially, in the TG regime of strong repulsive interactions, here for  $B<47.35$ G, losses are greatly suppressed, but increase rapidly on the attractive side of the CIR.  $N$ drops to a minimum when $B$ is increased and then recovers somewhat. A clear shift of the loss signature to lower values for $B$ and hence lower values for $a_\text{3D}$ can be discerned when the confinement is stiffened. When we identify the position of the edge with the position of the CIR, we find good agreement with the analytical result $C a_\text{3D} = a_{\perp}$ as shown in Fig.~\ref{fig2}(b). As we have no theoretical description of the detailed shape of the loss resonance, we also plot, for comparison, the position of the minimum, which is shifted accordingly.

In Fig.~\ref{fig2}(c) we juxtapose the loss and the heating rate that we measure in the vicinity of the CIR. For this, we measure the increase of the release energy within the first $100$ ms. After holding the atoms for time $\tau$ at a given value of $B$, we decrease $a_{\text{3D}}$ back to $250$ $a_0$ in $20$ ms, switch off the lattice potential and determine the release energy in the direction of the tubes from the momentum distribution in free space expansion. We observe an increase for the heating rate when the CIR is crossed. From a low value of $10$ nK/s in the TG regime it rises to a maximum of approximately $150$ nK/s and then drops to settle at some intermediate value. The position of the maximum agrees well with the maximum for atom loss. We check that the system's increase in energy is sufficiently small so that its 1D character is not lost. The release energy, even at maximal heating, remains below $k_B \times 30$ nK, which is far below the energy
spacing of the harmonic oscillator levels, $\hbar \omega_{\perp}\approx k_B \times 600$ nK.

\begin{figure}[t]
 \includegraphics[width=8.5cm] {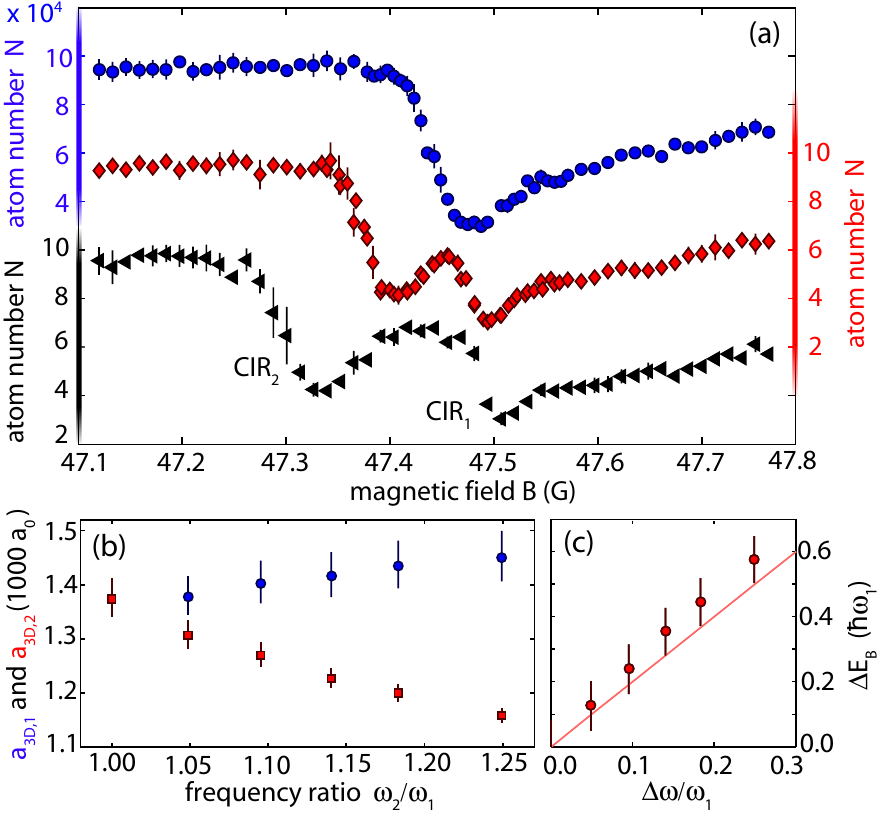}
 \caption{(color online) Splitting of a CIR for a 1D system with transversally anisotropic confinement. (a) As the horizontal confinement is stiffened, $\omega_2/\omega_1 = 1.00, 1.10, 1.18 $ (circles, diamonds, triangles)
 for $\omega_1=2 \pi \times 13.2(2)$ kHz, the CIR splits into CIR$_1$ and CIR$_2$. (b) Position of CIR$_1$ ($a_{\mathrm{3D},1}$, circles) and CIR$_2$ ($a_{\mathrm{3D},2}$, squares) as a function of the frequency ratio $\omega_2/\omega_1$. (c) Binding energy difference $\Delta E_\mathrm{B}$ as determined from the implicit equation (see text) in comparison to the expectation from the simple harmonic oscillator model (solid line).} \label{fig3}
\end{figure}

We now examine 1D systems with transversally anisotropic confinement. Starting from a lattice depth of $V=25\ E_R$ along both transversal directions, yielding $\omega_\perp = \omega_1 = \omega_2 = 2\pi \times 13.2(2)$ kHz, we increase the horizontal confinement to frequencies up to $\omega_2 = 2 \pi \times 16.5(2)$ kHz, corresponding to a lattice depth of $39\ E_R$, while keeping the depth of the vertical confinement constant. Fig.~\ref{fig3}(a) shows a distinct splitting of the original CIR into two loss resonances, CIR$_1$ and CIR$_2$. The splitting increases as the anisotropy is raised. In Fig.~\ref{fig3}(b) we plot the 3D scattering length values $a_{\mathrm{3D},1}$ and $a_{\mathrm{3D},2}$ that we associate with the positions of CIR$_1$ and CIR$_2$ as a function of the frequency ratio $\omega_2/\omega_1$. For this, as it becomes difficult to assign an edge to both of them, we simply determine the associated atom number minima and subtract a constant offset of $88(7) \ a_0$ as determined from the measurement shown in Fig.~\ref{fig2}(b). One of the resonances, CIR$_2$, exhibits a pronounced shift to smaller values for $a_\mathrm{3D}$ as the horizontal confinement is stiffened. The second resonance, CIR$_1$, shows a slight shift towards higher values for $a_\mathrm{3D}$. We now use the lifting of the degeneracy for the energy levels as indicated in Fig.~\ref{fig1}(b) to model the observed splitting of the CIR. We assume that the implicit equation $\zeta(1/2,-E_\mathrm{B}/(2\hbar\omega_{\perp}) + 1/2) = - a_{\perp}/a_{3D}$\, for the binding energy $E_\mathrm{B}$ \cite{Bergeman2003} remains approximately valid for sufficiently small $\Delta \omega$, taking $\omega_{\perp}= \omega_1$. Here, $\zeta$ is the Hurwitz zeta function. We translate the scattering length values $a_{\mathrm{3D},1}$ and $a_{\mathrm{3D},2}$ into binding energies and calculate the energy difference $\Delta E_\mathrm{B} = E_\mathrm{B}(a_\mathrm{3D,1}) - E_\mathrm{B}(a_\mathrm{3D,2})$, shown in Fig.~\ref{fig3}(c). While this model does not explain the upward deviation seen for CIR$_1$, the difference $\Delta E_\mathrm{B}$ is in reasonable agreement with the expected energy shift caused by the shifts of the excited harmonic oscillator states $(E_{0,2}-E_{2,0}) = 2\hbar \Delta \omega$ (solid line in Fig.~\ref{fig3}(c)). We thus attribute CIR$_2$ to the stiffened confinement along the horizontal direction and hence to state $(0,2)$, while CIR$_1$ corresponds to the unchanged confinement along the vertical direction and hence to state $(2,0)$.

\begin{figure}[t]
 \includegraphics[width=8.5cm] {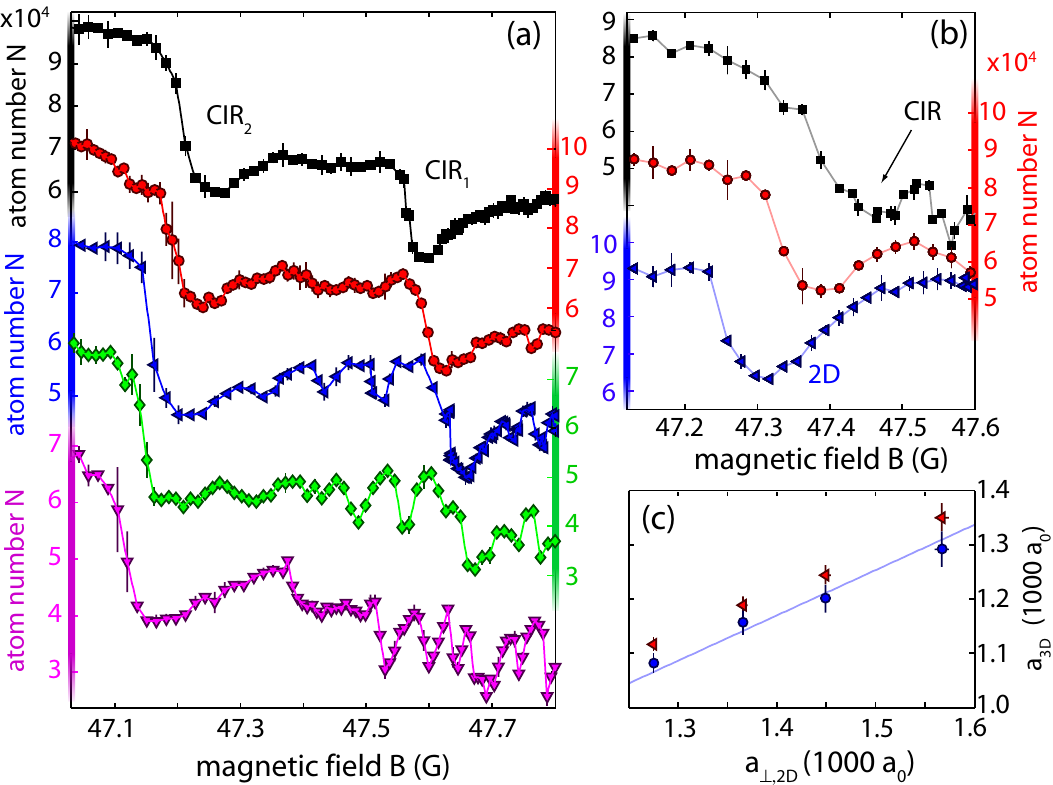}
 \caption{(color online) (a) Appearance of additional structure in the vicinity of CIRs for strongly anisotropic transversal confinement. The trap frequencies are $\omega_2=2\pi \times 16.6(2)$ kHz and $\omega_1/\omega_2 = 0.67, 0.60, 0.53, 0.49, 0.45 $ from top to bottom. (b) Transition from 1D to 2D confinement. As the horizontal lattice is ramped down, CIR$_2$ shifts and persists, while CIR$_1$ disappears ($\omega_1 = 2\pi \times 13.0(2)$ kHz and $\omega_2/\omega_1 = 0.58, 0.42, 0.00$ for squares, circles, and triangles). (c) Scaling of the CIR's position in 2D, in analogy to the 1D case shown in Fig.~\ref{fig2}(b). The position of the CIR as determined from the edge (circles) and, alternatively, from the minimum in atom number (triangles) shifts to lower values for $a_\text{3D}$ as the confinement is stiffened and $a_{\perp,\text{2D}}$ is reduced. The solid line is a linear fit according to $C_\text{2D} a_{\text{3D}}=a_{\perp,\text{2D}}$ with $C_\text{2D}=1.19(3)$.}
 \label{fig4}
\end{figure}

We observe the appearance of additional structure in the measured loss curves when we increase the transversal anisotropy by weakening the confinement along one axis, here along the vertical direction. Fig.~\ref{fig4}(a) shows the atom number after $\tau=300$ ms for trapping frequency ratios  $\omega_1/\omega_2$ from $0.67$ to $0.45$. Multiple loss resonances appear close to the position of CIR$_1$. The number of resonances increases and the positions shift continuously as the confinement is weakened. We speculate that those resonances are a result of a coupling to additional excited states, resulting in a multi-channel scattering situation. Also the weakening of the confinement could induce sufficient anharmonicity to allow for violation of the parity rule \cite{Peano2005}.

Surprisingly, we find that one of the CIRs persists in the limit of a 2D system. Previous theoretical studies on 2D systems have predicted the appearance of a CIR for negative $a_\text{3D}$, but not for positive $a_\text{3D}$ \cite{Petrov2001,Naidon2007}. In the experiment, we reduce the horizontal confinement while keeping the vertical confinement constant to probe the transition from the array of tubes to a stack of pancake-shaped, horizontally-oriented 2D systems. Trapping in the horizontal direction is still assured, now by the Gaussian profile of the vertically propagating laser beam, for which $\omega_2 = 2\pi\times 11$ Hz. Fig.~\ref{fig4}(b) shows that the CIR associated with the tight confinement shifts to lower values for $B$ and hence for $a_\text{3D}$ as the horizontal confinement is weakened. In the limit of 2D confinement, one of the CIRs, and in fact all the additional structure observed above, have disappeared, but one resonance persists. To check that the observed resonance is indeed the result of the 2D confinement, we vary the confinement along the tight vertical direction. Fig.~\ref{fig4}(c) plots the positions of edge and minimum of the loss signature as a function of $a_{\perp,\text{2D}}$, the confinement length associated with this direction. When we again associate the edge with the pole of the resonance, we obtain $ C_\text{2D} a_{\text{3D}}=a_{\perp,\text{2D}}$ with $C_\text{2D}=1.19(3)$, where $C_\text{2D}$ is a scaling factor similar to $C$ for the 1D case. Further scattering experiments are needed to elucidate the energy dependence of this 2D scattering resonance.

In summary, we have investigated the properties of CIRs, which appear in low-dimensional quantum systems as a result of tight confinement and which replace ``conventional'' 3D Feshbach resonances to tune the effective atomic interaction strength. We observed a splitting of the CIR for anisotropic transversal confinement, the appearance of multiple resonances for strongly anisotropic confinement, and the survival of one resonance for positive $ a_\text{3D} $ in the limit of 2D confinement. We expect that CIRs will not only be used in 1D geometry to tune the effective interaction strength as recently demonstrated \cite{Haller2009}, but also in 2D geometry and mixed dimensions \cite{Lamporesi2010} for the study of strongly-interacting quantum systems.

We thank W. Zwerger for discussions and R. Grimm for generous support. We acknowledge funding by the Austrian Ministry of Science and Research and the Austrian Science Fund and by the European Union within the framework of the EuroQUASAR collective research project QuDeGPM. R.H. is supported by a Marie Curie Fellowship within FP7. P.S. acknowledges financial support by the Deutsche Forschungsgemeinschaft. Financial support by the Heisenberg-Landau Program is appreciated by P.S. and V.S.M.

\bibliographystyle{apsrev}

\end{document}